\documentclass[letterpaper]{article} 
\usepackage{aaai25}  
\usepackage{times}  
\usepackage{helvet}  
\usepackage{courier}  
\usepackage[hyphens]{url}  
\usepackage{graphicx} 
\usepackage{natbib}  
\usepackage{caption} 
\usepackage{algorithm}
\usepackage{amsmath}
\usepackage{algorithmic}
\usepackage{array}
\usepackage{multirow}
\usepackage{newfloat}
\usepackage{listings}
\DeclareCaptionStyle{ruled}{labelfont=normalfont,labelsep=colon,strut=off} 
\floatstyle{ruled}
\newfloat{listing}{tb}{lst}{}
\floatname{listing}{Listing}
\title{DFT-Based Adversarial Attack Detection in MRI Brain Imaging: Enhancing Diagnostic Accuracy in
Alzheimer’s Case Studies}
\author{
Mohammad  Hossein Najafi\textsuperscript{\rm 1}, Mohammad Morsali\textsuperscript{\rm 1},\\ Mohammadmahdi Vahediahmar\textsuperscript{\rm 2}, Saeed Bagheri Shouraki\textsuperscript{\rm 1}
}
\affiliations{
    \textsuperscript{\rm 1}Department of Electrical Engineering, Sharif University of Technology\\
    \textsuperscript{\rm 2}Department of Computer Science, Drexel University\\

\{mh.najafi, mohammad.morseli\}@ee.sharif.edu, av834@drexel.edu, bagheri-s@sharif.edu
}

\begin{document}
\maketitle

\begin{abstract}
Recent advancements in deep learning, particularly in medical imaging, have significantly propelled the
progress of healthcare systems. However, examining the robustness of medical images against adversarial
attacks is crucial due to their real-world applications and profound impact on individuals’ health. These
attacks can result in misclassifications in disease diagnosis, potentially leading to severe consequences.
Numerous studies have explored both the implementation of adversarial attacks on medical images and
the development of defense mechanisms against these threats, highlighting the vulnerabilities of deep
neural networks to such adversarial activities.
In this study, we investigate adversarial attacks on images associated with Alzheimer’s disease and
propose a defensive method to counteract these attacks. Specifically, we examine adversarial attacks that
employ frequency domain transformations on Alzheimer’s disease images, along with other well-known
adversarial attacks. Our approach utilizes a convolutional neural network (CNN)-based autoencoder
architecture in conjunction with the two-dimensional Fourier transform of images for detection purposes.
The simulation results demonstrate that our detection and defense mechanism effectively mitigates several
adversarial attacks, thereby enhancing the robustness of deep neural networks against such vulnerabilities.
\end{abstract}

\section*{Introduction}
Deep Neural Networks have demonstrated remarkable capabilities across various real-world applications \cite{GU2018354}. In recent years, Convolutional Neural Networks (CNNs) have been extensively used in computer vision tasks. These networks are highly effective in facilitating feature extraction and pattern recognition, particularly in classification \cite{10.1007/s10462-022-10213-5}. However, recent studies have highlighted a critical issue: adversarial attacks significantly threaten their robustness. By introducing minimal perturbations to input images that are invisible to the human eye, these attacks can easily cause the networks to misclassify data, potentially undermining their reliability and accuracy in critical scenarios \cite{szegedy2014intriguingpropertiesneuralnetworks}.

DNNs play a critical role in medicine, especially in diagnosing various diseases \cite{LITJENS201760}. By utilizing MRI imaging, these networks can offer accurate and efficient analyses, greatly improving the precision of medical evaluations and treatment strategies.

Adversarial attacks on medical imaging and diagnosis carry severe potential consequences. Misdiagnoses caused by these attacks could lead to significant repercussions in the healthcare system and raise substantial security concerns \cite{finlayson2019adversarialattacksmedicaldeep}. For example, tampering with medical images could alter forensic reports and insurance claims \cite{10518037}. Furthermore, diagnostic errors could lead to inappropriate treatment prescriptions, posing serious risks to patient health.

Extensive research on adversarial attacks focuses on understanding these threats, developing countermeasures, and enhancing detection techniques. Significant progress in this area shows promise for better protection. In medical image analysis, deep neural networks (DNN) have shown promise in processing Alzheimer's disease (AD) images, enabling early and reliable detection, particularly in preclinical stages. Accurate assessment of disease progression helps clinicians predict patient outcomes, improving survival rates. Given Alzheimer's chronic nature, deep learning models can effectively analyze longitudinal time series data, capturing subtle feature changes crucial for early detection \cite{ElSappagh2023TrustworthyAI}.

Research on adversarial attacks and defenses in CNN models primarily targets natural images. Still, these methods may not translate well to medical images due to their unique characteristics and need for precision. Tailored strategies are essential for ensuring robust performance in medical imaging, underscoring the need for specialized research in this area \cite{10518037}.

The specific frequency components of images significantly affect adversarial perturbations \cite{9534307}. Deep neural networks (DNNs) used for image classification are sensitive to various frequency information. Studies suggest that classification models rely on texture information in high-frequency components and shape information in low-frequency image components.

\begin{figure*}[t]
\centering
\includegraphics[width=1\textwidth]{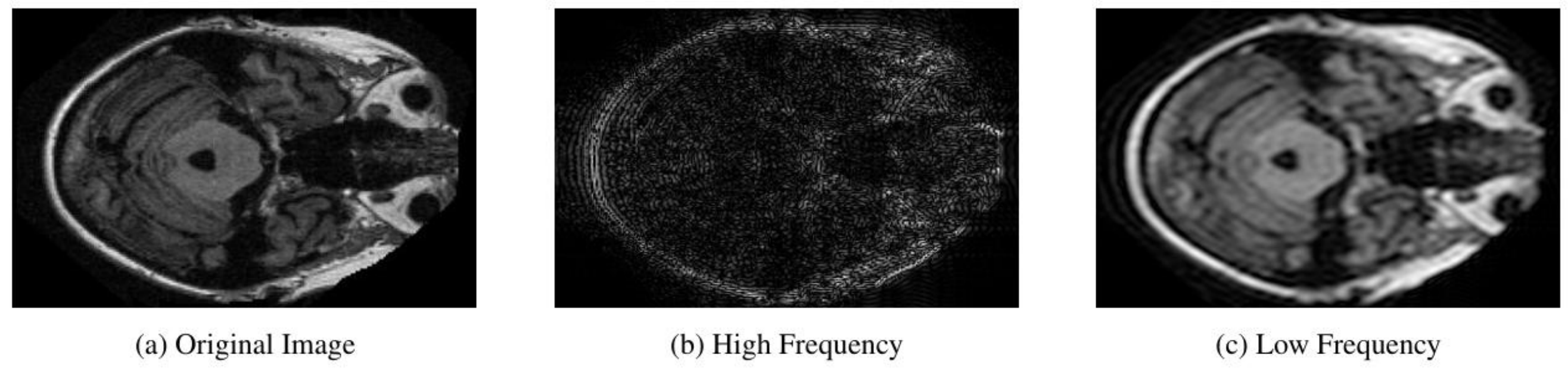} 
\caption{Comparison of Original, High Frequency, and Low Frequency Images. Image is from Kaggle OASIS dataset \cite{daithal_imagesoasis_2024}.}
\label{fig1}
\end{figure*}

Current frequency-based attack algorithms attempt to deceive models by searching for perturbation patterns across the entire frequency spectrum \cite{10.1007/978-3-030-68238-5_36,10.1007/978-3-031-19772-7_32}. However, these algorithms tend to produce high-frequency solutions. On the other hand, empirical studies have demonstrated that low-frequency perturbations are more effective in misleading image classification models. Restricting adversarial perturbations to specific frequency components of images can significantly improve their impact. Deep neural networks (DNNs) for image classification are sensitive to different frequency information. Research indicates that classification models exploit texture information in the high-frequency components and shape information in the low-frequency components of images. Figure \ref{fig1} shows a comparison of original, high-frequency, and low-frequency Alzheimer's disease Images.

We analyze the contribution of various image components in the frequency domain to the model's inference outcomes. Research limits the search space for adversarial perturbations to the element most significantly influencing these outcomes. Wavelet-based decomposition is an optimal method for producing an image's different frequency domain components. This approach is highly effective because an image's approximate composition encompasses features that nearly all models rely on \cite{10.1007/978-3-030-66415-2_10,Sarvar2023,song2024wpdafrequencybasedbackdoorattack}.

Therefore, examining and defending against frequency domain-based adversarial attacks alongside well-known and basic attacks is important. In the following sections, we will detail the types of adversarial attacks and the methods we have employed to counter and detect them. Specifically, we examine various frequency domain-based adversarial attacks, then introduce our defense strategy, which leverages U-Net-based autoencoders and Fourier transforms, along with relevant loss functions, to enhance Alzheimer's disease medical imaging \cite{oktay2018attentionunetlearninglook, Fan2021, CHEN2023107248}.

\section*{Related Works}

Initially, we introduce attack algorithms targeting our models for Alzheimer's disease image classification, followed by those relevant to our research.

In 2014, Goodfellow, \cite{Goodfellow2014ExplainingAH} and his colleagues demonstrated that it is possible to generate inputs that a neural network cannot classify correctly by making small perturbations to the input of convolutional neural networks using a method similar to gradient descent. Since then, numerous new attack methods have been introduced in recent years.

For example, Nir Morgulis and his colleagues \cite{Morgulis2019FoolingAR} have demonstrated that creating adversarial examples can trick Tesla's self-driving systems, leading to the vehicle executing incorrect commands. The potential for such attacks to result in disastrous consequences is clear.

\subsection*{1. Adversarial Attacks }

\subsubsection*{ Fast Gradient Sign Method (FGSM)}

This method generates adversarial examples by moving in the direction opposite to the gradient \cite{madry2019deeplearningmodelsresistant}. The cost function of the optimized network model is represented by \( J = (\theta, x, y) \), where \( \theta \) denotes the correct input, \( x \) the clean input, and \( y \) the network output.

For an untargeted attack, the adversarial example is:

\[
x_{\text{adv}} = x + \epsilon \cdot \text{sign}(\nabla_x J)
\]

Here, \( \epsilon \) is the noise magnitude added to the image. The clean input \( x \) moves \( \epsilon \) in the gradient direction, making the adversarial example nearly indistinguishable from the original input.

\subsubsection*{ Basic Iterative Method (BIM)}
In 2016, to improve upon the FGSM method, Kurakin and his colleagues introduced the Basic Iterative Method (BIM) \cite{10.1007/978-3-319-94042-7_11}, an iterative approach for creating adversarial examples. In BIM, FGSM attacks are applied repeatedly with small step sizes. These iterative updates continue until the accumulated disturbances surpass a specific threshold a human observer notices.

The adversarial example at each iteration \( t \) is updated as follows:
\[ x_{\text{adv}}^{t+1} = \text{Clip}\left(x_{\text{adv}}^t + \alpha \cdot \text{sign}(\nabla_x J(\theta, x, y))\right) \]

The clip function ensures that the updated image remains within a valid range, considering the number of steps \( T \) in the BIM method. The algorithm is terminated if the total perturbation exceeds a certain limit \( \epsilon \).

\subsubsection*{ Projective Gradient Descent (PGD)}
The Projective Gradient Descent (PGD) attack is similar to the Basic Iterative Method (BIM) attack, employing an iterative approach based on the gradient of the cost function to create adversarial examples. However, the key distinction lies in the step size utilized; unlike the BIM attack, PGD does not terminate at a fixed step size \( \epsilon \) but continues iterating until the perturbation norm reaches a specified threshold \cite{madry2019deeplearningmodelsresistant}. Upon each iteration of the algorithm, the image is projected back onto the \( \epsilon \)-ball surrounding the original sample, ensuring that the perturbation remains within the defined bounds.
This attack is defined as follows:
 
\[
x_{\text{adv}}^{t+1} = \Pi_{x + \epsilon} \left( x_{\text{adv}}^t + \alpha \cdot \text{sign}(\nabla_x J(\theta, x, y)) \right)
\]
In this equation, the function \( \Pi_{x + \epsilon}(x) \) ensures that the image remains within the \( \epsilon \)-ball around the original image at each iteration.
\\
Current research focuses primarily on creating adversarial examples with spatial constraints, which sometimes results in significant degradation of image quality and perceptible perturbations to human observers. Drawing insights from recent studies in attacking classifiers in the frequency domain \cite{10.1007/978-3-030-01264-9_28,8954086,wang2020highfrequencycomponenthelps,Wang2020TowardsFE,9878492}
, we leverage low-frequency components extracted from decomposed input samples to constrain perturbations more precisely.

Recent studies have explored how deep neural networks (DNNs) generalize and their vulnerability to adversarial attacks from a frequency perspective. Initial studies suggest that DNNs can capture high-frequency elements that humans do not easily perceive. Yin showed that naturally trained models are susceptible to perturbations in high-frequency components, while adversarially trained models are less sensitive to such perturbations \cite{yin2020fourierperspectivemodelrobustness} . As a result, various methods have been suggested to create adversarial examples by considering frequency perspectives. For instance, Long introduces a technique where input images are altered with Gaussian noise in the frequency domain as a form of data augmentation and then converted back to the spatial domain for gradient calculation, thus improving transferability \cite{long2022frequencydomainmodelaugmentation}. Guo introduced an attack involving a random search in the low-frequency band of the DCT domain \cite{guo2019lowfrequencyadversarialperturbation}. Sharma found that image classification models are not robust against low-frequency attacks \cite{sharma2019effectivenesslowfrequencyperturbations}. And Wang Crafted Transferable Targeted Adversarial Examples with Low-Frequency Perturbations, which train a conditional generator to generate targeted adversarial perturbations that are then added to the low-frequency component of the image \cite{wang2023lfaacraftingtransferabletargeted}. Yang sought to explore a feature contrastive approach in the frequency domain to generate impactful adversarial examples in both cross-domain and cross-model settings \cite{Yang_Jeong_Yoon_2024}. These existing approaches overly concentrate on the low-frequency component of the image, and this approach is largely based on intuition and experience.

Also wavelets are widely used in signal processing and pattern recognition \cite{192463}. In deep learning, the wavelet transform is commonly used for image pre-processing or post-processing. The discrete wavelet transform (DWT) breaks an image into four components by sampling its horizontal and vertical channels using sub-band filters. These components correspond to the image's approximation, vertical, horizontal, and diagonal details. The wavelet transform's ability to localize time-frequency provides an advantage over the Fourier or cosine transforms in image processing. The inverse discrete wavelet transform (IDWT) can reconstruct the original data from the DWT output \cite{8237449,8954076}.

Adversarial attacks often introduce noise in the spatial domain. Each frequency component carries different levels of importance in its spatial representation in the frequency domain. Our goal is to identify the frequency band that significantly influences model inference by analyzing the effects of different frequency bands on the model in that region. This approach aims to develop adversarial noise that generalizes well across various models \cite{anshumaan2020wavetransformcraftingadversarialexamples,https://doi.org/10.1002/int.23031}.

\subsubsection*{ Discrete Wavelet Transform Fast Gradient Sign Method (DWT-FGSM) \cite{anshumaan2020wavetransformcraftingadversarialexamples}}
The Discrete Wavelet Transform Fast Gradient Sign Method (DWT-FGSM) enhances the traditional FGSM attack by generating adversarial examples in the frequency domain using wavelet transforms. It targets low-frequency image components, which capture general structure and smooth variations while preserving high-frequency details like edges.

\subsubsection*{ Discrete Wavelet Transform Projective Gradient Descent (DWT-PGD) \cite{anshumaan2020wavetransformcraftingadversarialexamples}}
The Discrete Wavelet Transform Projective Gradient Descent (DWT-PGD) method extends the PGD attack by operating in the frequency domain using wavelet transforms. This approach allows for precise manipulation of an image's frequency components, particularly targeting low-frequency elements. By decomposing the image into low-frequency and high-frequency components, the method focuses on iteratively updating the low-frequency component while keeping perturbations within allowed limits. The adversarial image is reconstructed, combining the modified low-frequency component with the original high-frequency details.

\subsubsection*{ Discrete Wavelet Transform Auto Projective Gradient Descent (DWT-Auto-PGD) \cite{anshumaan2020wavetransformcraftingadversarialexamples,croce2020reliableevaluationadversarialrobustness,lorenz2024detectingautoattackperturbationsfrequency}} 
The Discrete Wavelet Transform Auto Projective Gradient Descent (DWT-Auto-PGD) method is an advanced version of PGD that optimizes performance by dynamically adjusting step sizes during the attack process. By operating in the frequency domain using DWT, this method focuses on low-frequency components to generate adversarial examples. The wavelet transform's ability to localize spatial and frequency information effectively exploits neural network vulnerabilities. DWT-Auto-PGD iteratively updates the low-frequency component of an image, adjusting the step size based on the loss to ensure optimal adversarial perturbation within the allowed limits while maintaining minimal perceptibility.

\subsubsection*{ Spectrum Simulation Attack \cite{long2022frequencydomainmodelaugmentation}}
The frequency attack method presented by  Long et al. generates adversarial examples by manipulating the frequency components of input data. This approach transforms the input data into the frequency domain using the Discrete Cosine Transform (DCT), allowing targeted perturbations. Specifically, perturbations are applied to mid-frequency components, balancing perceptibility and effectiveness. These perturbations are optimized to be subtle yet impactful. The altered data is then converted back to the spatial domain using the Inverse Discrete Cosine Transform (IDCT). This method leverages frequency domain properties to create efficient and effective adversarial attacks. 
\subsection*{2. Defense Methods}
\begin{figure}
\includegraphics[width=0.48\textwidth]{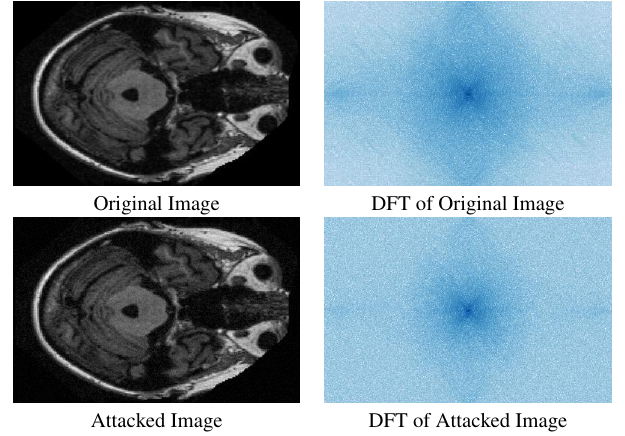} 
\caption{Comparison of Original and Attacked Alzheimer's Disease MRI Images and Their DFT. Image is from Kaggle OASIS dataset.}
\label{fig2}
\end{figure}

In frequency-based adversarial defense, Zhang removed high-frequency noise in the Fourier space to reduce the impact of adversarial perturbations \cite{Zhang2019AdversarialDB}. Similarly, Huang applied regularization to the input image in the Fourier space to remove adversarial noise \cite{huang2022frequencyregularizationimprovingadversarial}. A defense method introduced by Sitawarin involves using the K-Nearest Neighbor (KNN) algorithm to detect adversarial data \cite{Sitawarin2019DefendingAA}. In this technique, the input data is first grouped using KNN, and then the Euclidean distance of each unknown input to the cluster centers is measured. If this distance exceeds a predefined margin, the input is identified as adversarial and is blocked from entering the network. Expanding on Sitawarin’s approach, Harder developed a Fourier-based defense that clusters the input dataset in the frequency domain using the K-Nearest Neighbor algorithm and then assesses the adversarial nature of an unknown input based on its distance to the clusters \cite{Harder2021SpectralDefenseDA}. Then, Shah introduced frequency-centric defense mechanisms against adversarial examples by training an adversarial detector and denoising the adversarial effect \cite{10.1145/3475724.3483610}. After that, Li developed a robust attention ranking architecture with frequency-domain transform to defend against adversarial samples called RARFTA \cite{LI2023103717}. They imported the discrete cosine transform as the activation layer after the first convolutional layer, which effectively suppresses the attack based on the gradient method. \\
We utilize the concept of a manifold to measure the Euclidean distance. Zhou employed the manifold to project samples onto the input \cite{10.1007/978-3-030-58577-8_18}. Our objective is to design a detector that initially projects the input dataset onto a suitable manifold and then determines whether the input image is clean or adversarial by measuring the Euclidean distance. This approach was first introduced by Meng \cite{meng2017magnettwoprongeddefenseadversarial}. This approach estimates a suitable manifold using an autoencoder and determines a threshold (\(t_{re}\)) based on the Euclidean distance from the manifold. If a sample's distance exceeds this threshold, it is classified as adversarial. The threshold \( t_m \) is a key hyperparameter, requiring careful selection to minimize false positives and negatives. A separate validation set helps determine this threshold, ensuring clean data isn't misclassified. While the method proposed by Meng et al. performs well on various datasets without needing knowledge of the network or attack type, it struggles with sparse datasets, like medical data, where small Euclidean distances between classes can lead to misclassification \cite{lu2018limitationmagnetdefensel1based}.
\\
Our idea, implemented in the present research, involves using the frequency domain to measure the Euclidean distance relative to the manifold (Figure \ref{fig2}). Various methods have discussed using a Fourier transform to convert the input image to remove or identify adversarial noise. However, we use the input dataset's manifold in the frequency domain to calculate the Euclidean distance \cite{Maiya2021AFP,Zhang2019AdversarialDB,Harder2021SpectralDefenseDA,raviv2021enhancingrobustnessneuralnetworks,hemmati2023adversarial}.

\begin{figure*}[t]
\centering
\includegraphics[width=0.9\textwidth]{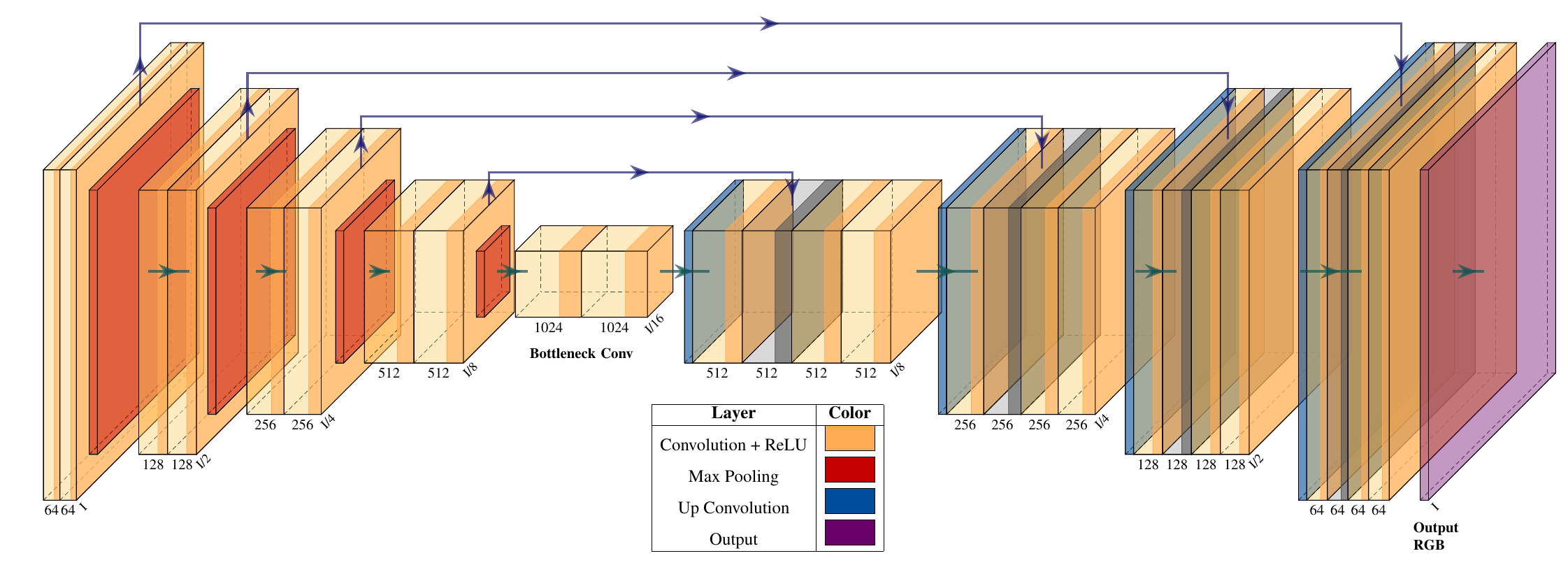} 
\caption{Architecture of U-Net-Based Autoencoder}
\label{fig3}
\end{figure*}
Robust defense methods have recently involved designing networks that project the input dataset onto an appropriate manifold, often using an autoencoder. The bottleneck layer of the autoencoder captures the dataset with reduced dimensions, and the decoder reconstructs the images. The encoder projects the input onto the manifold while the decoder handles reconstruction. U-Net, a CNN architecture introduced by Ronneberger et al. in 2015, is particularly effective for image segmentation, especially in medical image analysis, by accurately classifying each pixel into different regions or classes \cite{9446143}. The U-Net architecture consists of two main parts: the contracting and expansive paths. The contracting path functions similarly to a typical CNN encoder, while the expansive path acts as a decoder. \\The U-Net architecture is widely used in various medical imaging applications, such as segmenting organs, tumors, or other structures of interest in CT or MRI scans. Its ability to effectively capture local and global features through skip connections has made it a popular choice for tasks requiring precise image segmentation. By leveraging the encoder-decoder structure of U-Net as an autoencoder, we can modify and reconstruct the input image. Figure \ref{fig3} shows architecture of U-Net-Based Autoencoder. The use of U-Net as an autoencoder is described below \cite{9057834,haris_iqbal_2018_2526396}:

\begin{itemize}
    \item \textbf{Encoder:} The encoder part of the U-Net architecture remains unchanged. 
    
    \item \textbf{Bottleneck:} The main difference in using U-Net for segmentation is the existence of a bottleneck layer, where the spatial dimensions of the feature maps are minimized. This layer can be adjusted to maintain higher spatial resolution in the autoencoder version, allowing for better reconstruction. This adjustment can involve modifying the number of filters or reconfiguring the layer. This layer essentially represents the manifold fitted to the dataset.

    \item \textbf{Decoder:} The decoder part of the U-Net architecture must be modified to facilitate the reconstruction of the corrected image. Instead of upsampling operations and concatenating to restore the primary image resolution, transposed convolutions (deconvolutions) can be used. Transpose convolutions are capable of remapping the features. These transpose convolutions and upsampling layers learn the filters needed for remapping the features.
    \item \textbf{Output Layer:} In the main U-Net used for segmentation, the output layer usually employs a 1x1 convolution with a sigmoid or softmax activation function to generate pixel-wise segmentation maps. For an autoencoder, however, the output layer should use an appropriate activation function (like ReLU) for reconstructing the image, and the number of output channels should match the number of input channels (e.g., three channels for an RGB image).

    \item \textbf{Loss Function:} The loss function most commonly used for training an autoencoder generally involves comparing the reconstructed image with the original input image. Common choices include Mean Squared Error (MSE) or Mean Absolute Error (MAE), which measure the pixel-wise difference between the input and the reconstructed image.

\end{itemize}

Moreover, we employed various loss functions in addition to the mean squared error for training the autoencoder. These loss functions are predominantly used in medical imaging and related domains. The MONAI library uses these loss functions based on well-established research. 

\subsection*{3. Loss Functions for Training Autoencoder \cite{cardoso2022monaiopensourceframeworkdeep}}

\begin{itemize}
    \item \textbf{Sure Loss: } Calculate Stein’s Unbiased Risk Estimator (SURE) loss for a given operator.

This differentiable loss function can train/guide an operator (e.g., neural network) where the pseudo ground truth is available but the reference ground truth is not. For example, in the MRI reconstruction, the pseudo ground truth is the zero-filled reconstruction, and the reference ground truth is the fully sampled reconstruction. The reference ground truth is often unavailable due to the lack of fully sampled data.

The original SURE loss is proposed in \citealp{Stein1981EstimationOT}, and the SURE loss used to guide the diffusion model-based MRI reconstruction is proposed in \citealp{10.1007/978-3-031-43898-1_20}.

    \item \textbf{Diffusion Loss: }Calculate the diffusion based on first-order differentiation of prediction using central finite difference \cite{Balakrishnan_2019}.

The diffusion loss function in image registration ensures the smoothness of the deformation field, which aligns one image to another. It penalizes rapid or irregular changes in the displacement vector field, ensuring that the deformations are physically plausible and anatomically consistent. This loss is calculated by considering the gradients of the displacement field across all spatial dimensions using the central finite difference method. The resulting loss function is integrated into the model's training and is balanced by a regularization parameter to ensure accurate image alignment and smooth deformation. This approach improves the reliability of registered images for clinical use. 
\end{itemize}

In this study, we aim to develop a robust method to detect adversarial attacks in MRI brain images, particularly focusing on Alzheimer's disease cases. Our approach leverages the Discrete Fourier Transform (DFT) and an autoencoder architecture to distinguish between normal and adversarially manipulated images.

Adversarial attacks are characterized by subtle perturbations that cause deep neural networks to make incorrect predictions despite the changes being invisible to the human eye. This is largely due to the high dimensionality of deep networks and the sparse distribution of data within these dimensions, which increases the potential for misclassification. In contrast, the human eye is resilient to these attacks because it operates in a lower-dimensional space and utilizes prior knowledge for classification and recognition.

To address deep networks' vulnerability to adversarial attacks, we propose a method that reduces the dimensionality of image recognition. By transforming images into the frequency domain using DFT and employing an autoencoder trained on clean data, we can potentially detect adversarial examples by analyzing the reconstruction errors.

We utilized a dataset of MRI brain images comprising normal and Alzheimer's disease-affected scans. To create adversarial examples, we employed several known adversarial attack techniques. These attacks were designed to subtly modify the original images in a way that could mislead deep neural network classifiers.

\subsection{Feature Extraction Using Discrete Fourier Transform (DFT)}
Our method for robustness is based on detecting adversarial examples and
The core of our detection method relies on the images' frequency domain characteristics. We applied the two-dimensional Discrete Fourier Transform (2D-DFT) to normal and adversarial images. The DFT transforms the images' spatial representation into the frequency domain, highlighting differences that may not be evident in the spatial domain. The human eye is sensitive to changes in the frequency domain, making it a valuable feature for detecting adversarial perturbations. 

Before being given to the classification model, the input images are processed through an attack detection module. This module utilizes three distinct methods to identify tampered images. If an image is detected as tampered, it is not forwarded to the classification model. Conversely, if the image is considered intact, it is passed on to the classification model for further processing. 
\section{Methodology}

\subsection{Autoencoder Architecture and Training}
We developed a CNN-based autoencoder to learn the distribution of clean images. The encoder uses convolutional and pooling layers to capture essential features, and the latent space helps distinguish between normal and adversarial images, as perturbations can cause significant deviations here. The decoder, mirroring the encoder, reconstructs the image using convolutional and upsampling layers. Trained solely on clean MRI images, the autoencoder minimizes reconstruction error to recreate images from latent representations accurately.

\subsection{Detection Mechanism}

For adversarial attack detection, we applied the 2D-DFT to convert the input image to the frequency domain and then passed it through the trained autoencoder to obtain a latent representation. The decoder then reconstructed the image from this representation. The reconstruction error between the input and output images was analyzed, as adversarial images typically yield higher errors due to perturbations. The latent space was also visualized as a data manifold, with clean data residing within a defined Euclidean distance region; points outside this region were identified as potential adversarial examples. This method also enhances detection when applied to the autoencoder's output.

To evaluate the effectiveness of our detection method, we conducted simulations. We measured the reconstruction errors for both normal and adversarial images, establishing a threshold to differentiate between the two. The performance accuracy was calculated to quantify the robustness of our method against various adversarial attacks.
\section{Experiment}

\begin{table*}[t]
\centering
\setlength{\tabcolsep}{1.5mm} 
\renewcommand{\arraystretch}{1.3} 
\begin{tabular}{|>{\centering\arraybackslash}m{1.9cm}|>{\centering\arraybackslash}m{2.15cm}|>{\centering}m{1.33cm}|>{\centering}m{1.33cm}|>{\centering}m{1.33cm}|>{\centering}m{1.33cm}|>{\centering}m{1.33cm}|>{\centering\arraybackslash}m{1.33cm}|}

\hline
\textbf{Model} & \textbf{Loss Function} & \multicolumn{2}{c|}{\textbf{Loss}} & \multicolumn{2}{c|}{\textbf{Decoded}} & \multicolumn{2}{c|}{\textbf{Encoded}} \\
\cline{3-8}
& & \textbf{$\epsilon$ = 4/255} & \textbf{$\epsilon$ = 8/255} & \textbf{$\epsilon$ = 4/255} & \textbf{$\epsilon$ = 8/255} & \textbf{$\epsilon$ = 4/255} & \textbf{$\epsilon$ = 8/255} \\
\hline
\textbf{} & Diffusion  & \textbf{97.95} & 96.59 & 42.19 & 62.37 & 41.86 & 64.08 \\
\textbf{Simple CNN}  & Sure  & 97.56 & 96.39 & 42.43 & \textbf{62.96} & \textbf{65.14} & \textbf{64.44} \\
 & MSE  & 97.49 & \textbf{97.04} & \textbf{44.83} & 62.49 & 42.95 & 61.69 \\
\hline
\textbf{} & Diffusion  & 85.8 & \textbf{87.76} & 54.4 & 62.86 & 53.58 & 61.2 \\
\textbf{Inception v3}& Sure  & 85.58 & 87.25 & \textbf{55.02} & \textbf{63.12} & \textbf{64.04} & \textbf{64.92} \\
 & MSE  & \textbf{86.02} & 87.3 & 54.93 & 54.89 & 54.34 & 56.12 \\
\hline
\textbf{} & Diffusion  & 90.58 & \textbf{91.54} & 52.97 & 68.12 & 52.51 & 67.76 \\
 \textbf{ResNet-50} & Sure  & 91.04 & 91.48 & 51.89 & 68.55 & \textbf{73.75} & \textbf{68.67} \\
 & MSE  & \textbf{91.17} & 91.18 & \textbf{53.32} & \textbf{68.58} & 53.17 & 62 \\
\hline
\textbf{} & Diffusion  & 86.61 & \textbf{86.65} & 61.17 & 63.68 & 61.18 & 50.05 \\
\textbf{VGG16} & Sure  & \textbf{87.78} & 85.75 & \textbf{61.75} & \textbf{65.39} & \textbf{80.36} & \textbf{66.63} \\
 & MSE  & 78.87 & 85.69 & 59.65 & 64.81 & 58.98 & 52.31 \\
\hline

\end{tabular}
\caption{Model performance under different conditions. The results are reported for four models (Simple CNN, Inception v3,  Resnet-50, VGG16) with three different loss functions (Diffusion, Sure, MSE). The performance is measured in terms of accuracy under adversarial attacks with different epsilon values, detection methods (Loss, Decoded, Encoded).}
\label{table2}
\end{table*}

\begin{table}[t]
\centering
\setlength{\tabcolsep}{1.5mm} 
\renewcommand{\arraystretch}{1.3} 
\begin{tabular}{|>{\centering\arraybackslash}p{2cm}|>{\centering}m{1.8cm}|>{\centering}m{1.8cm}|>{\centering\arraybackslash}m{1.2cm}|}

\hline
\textbf{Model} & \multicolumn{2}{c|}{\textbf{Attack}} & \textbf{Clean} \\
\cline{2-3}
& \textbf{$\epsilon$ = 4/255} & \textbf{$\epsilon$ = 8/255} & \\
\hline
\textbf{\centering Simple CNN} & 41.87 & 40.09 & 97.95 \\
\textbf{\centering Inception v3} & 52.14 & 48.76 & 91.33 \\
\textbf{\centering ResNet-50} & 51.57 & 47.05 & 94.7 \\
\textbf{\centering VGG16} & 61.75 & 44.25 & 91.08 \\

\hline

\end{tabular}
\caption{Model performance under different conditions. The results are reported for four models (Simple CNN, Inception v3, ResNet-50, VGG16) under adversarial attacks with different epsilon values (4/255, 8/255) and on clean data.}
\label{table1}
\end{table}

We used the Kaggle OASIS MRI brain imaging dataset, which includes 86,437 samples across the following four classes: Non-Demented (67,222 samples), Mild Dementia (5,002 samples), Moderate Dementia (488 samples), and Very Mild Dementia (13,725 samples). The sample distribution across these classes is imbalanced, which could affect the learning process and result in poor performance. Techniques such as downsampling and upsampling were applied to address this issue, ensuring a balanced distribution for training process.

\subsection{Models}

\begin{figure}
\includegraphics[width=0.55\textwidth]{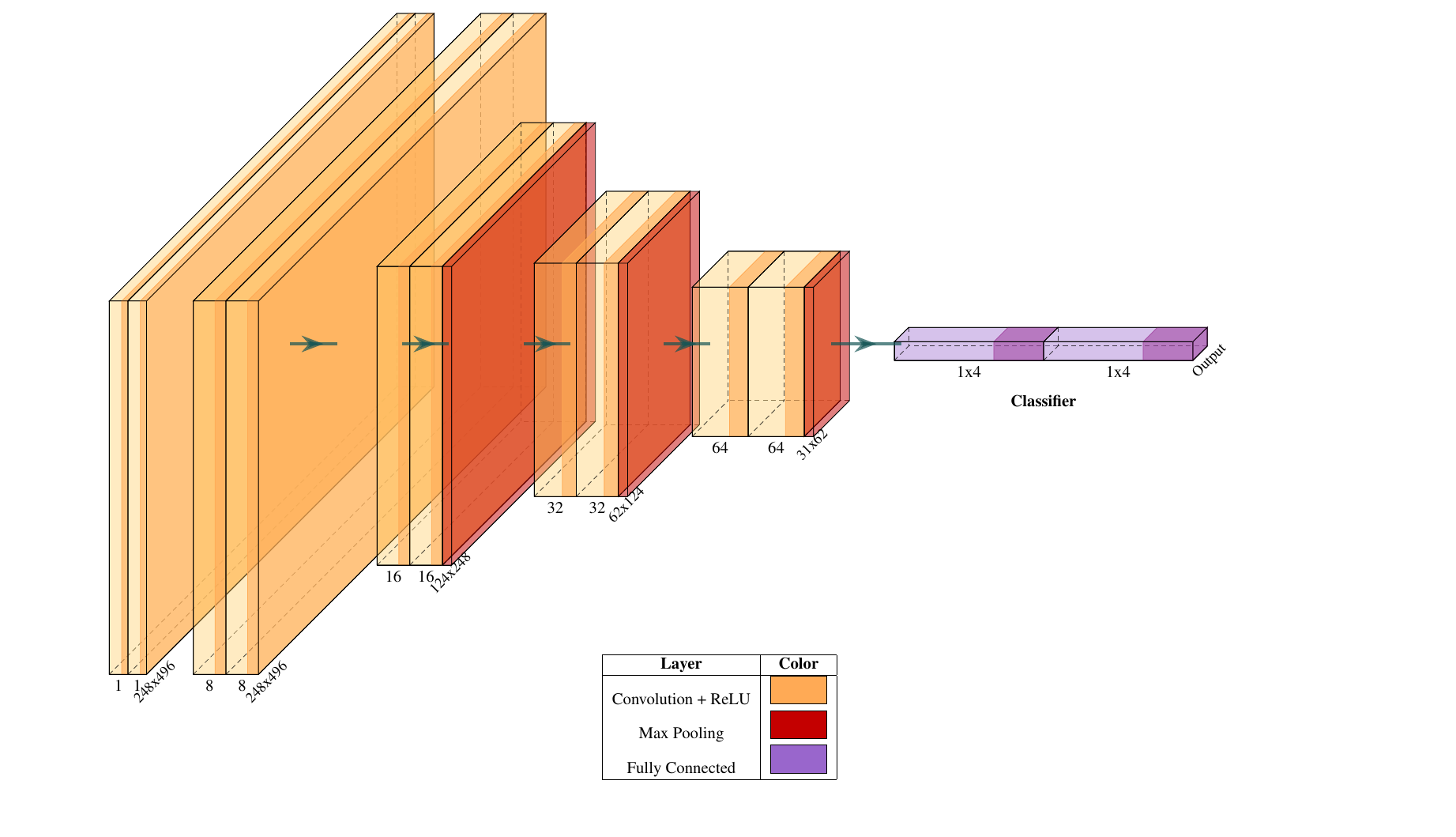} 
\caption{Architecture of Simple CNN}
\label{fig4}
\end{figure}

We employed advanced deep neural network models for image classification, known for their excellent performance in learning and adaptability across various scenarios due to their ability to capture intricate patterns. Specifically, we worked with the Inception v3, VGG16, and ResNet-50 models \cite{simonyan2015deepconvolutionalnetworkslargescale,he2015deepresiduallearningimage,szegedy2015rethinkinginceptionarchitecturecomputer}. We also used a simpler model with fewer layers, called Simple CNN, mainly for comparison (Figure \ref{fig4}). In this simple model, we provided three-channel images as input to this network instead of the single-channel images used instead of the original model \cite{wong_shallow_2024}.

\subsection{Attacks}
We utilized well-known Gradient-Based attacks such as FGSM, PGS, and BIM, as well as frequency-based attacks, including DWT-FGSM, DWT-PGD, DWT-Auto-PGD, and Spectrum. At this stage, the dataset related to the testing phase is divided. 37\% of the data remains clean, while the remaining 63\% is divided equally among these seven types of attacks.

\subsection{Autoencoder}
We employed a U-Net-based autoencoder, shown in Figure \ref{fig3}, and trained it on the entire dataset with 15 epochs. This autoencoder's learning process is focused solely on reconstruction, not classification.
We utilized different loss functions to train the autoencoder: MSE, Diffusion, and Sure. However, it is important to note that this autoencoder does not use the image itself, and the input to the autoencoder is the DFT of the images.
\subsection{Implementation Details
}
During the learning phase, 80\% of the dataset was used, and the remaining 20\% was set aside for testing. For each model in Table \ref{table1}, trained with 30 epochs. The classification accuracy for each model is listed under the "clean" section of Table \ref{table1}, with results corresponding to attacks with epsilon values of 4/255 and 8/255. The attack detection methods were used to determine whether an attack had occurred and prevent it from entering the model. The results of this detection process using the three detection methods and three loss functions for each model are presented in Table \ref{table2}.

\section{Conclusion and Future Works}

Our methodology, which combines DFT with a CNN-based autoencoder, has significant practical implications. It provides a robust framework for detecting adversarial attacks in MRI brain images, thereby enhancing the security and reliability of deep learning models in medical imaging applications. By focusing on frequency domain features and leveraging the autoencoder's ability to learn the distribution of clean images, our approach can be applied to improve the safety and accuracy of medical diagnoses.

In future work, we will explore other features beyond two-dimensional Fourier transforms of images and address the bias in accuracy when using multiple image features to counter adversarial attacks. Additionally, we will investigate features and methods for detecting adversarial attacks that aim to mimic the behavior of the human eye and brain.   

\section{Acknowledgment}
We sincerely thank Saman Soleimani, and Shaghayegh Vesali for their invaluable insights and thoughtful discussions, which helped develop this work.

\bibliography{aaai25}

\end{document}